\documentclass[%
 reprint,
 amsmath,amssymb,
 aps,
]{revtex4-2}
\usepackage{xcolor}
\usepackage{graphicx}
\usepackage{dcolumn}
\usepackage{bm}

\begin{document}

\preprint{APS/123-QED}

\title{Bootstrapping Fermionic Veneziano Amplitudes from Regge Zeros}

\author{Pongwit Srisangyingcharoen}
 \email{pongwits@nu.ac.th}
\affiliation{%
 The Institute for Fundamental Study, Naresuan University, Thailand
}%

\date{\today}

\begin{abstract}
We extend the Regge-zero bootstrap of the Veneziano amplitude to tree-level scattering with external
fermions. We consider the colour-ordered four-point amplitude of a massless scalar and a massless
spin-$\frac12$ fermion, in which fermionic resonances appear in the $s$-channel and bosonic resonances
in the $t$-channel. Factoring out a spinor structure that carries the little-group weight of the
amplitude, we bootstrap the resulting reduced amplitude, whose residues in each channel vanish at
Regge zeros controlled by the trajectory of the opposite channel. Assuming ultrasoft Regge behaviour
and that the residues have no zeros other than Regge zeros, we show that both Regge trajectories are
linear and that the maximal spins grow with the level as $J_F=n+\frac12$ and $J_B=m+1$. Duality
between the fermionic- and bosonic-pole expansions then fixes the reduced amplitude to a Beta
function, up to its normalisation and the ratio of the two Regge slopes. Requiring the fermionic and
bosonic spectra to be degenerate equates the slopes, and the resulting amplitude coincides with the
open superstring amplitude.
\end{abstract}

\maketitle


\section{Introduction}\label{sec:1}

The idea that consistency alone might determine the scattering of hadrons drove the $S$-matrix
programme of the 1960s. Regge's analysis of complex angular momentum~\cite{Regge:1959mz} showed that
high-energy scattering is governed by trajectories $\alpha(t)$ that interpolate between resonances of
increasing spin, and the approximately linear Chew--Frautschi plots of hadron
spectra~\cite{Chew:1961ev,Chew:1962eu} pointed to infinite towers of states. The finite-energy sum
rules of Dolen, Horn and Schmid~\cite{Dolen:1967jr} then suggested that the sum over $s$-channel
resonances and the sum over $t$-channel Regge exchanges are two descriptions of the same amplitude.
Veneziano's amplitude~\cite{Veneziano:1968yb}, followed by its closed-string counterpart due to
Virasoro and Shapiro~\cite{Virasoro:1969me,Shapiro:1970gy}, provided the first explicit realisations
of this duality, with linear trajectories and Regge behaviour built in. These dual resonance models
were soon recognised as scattering amplitudes of relativistic strings and became the starting point
of string theory~\cite{Green:1987sp,Polchinski:1998rq}. Reviews of these early developments can be
found in Refs.~\cite{Collins:1977jy,Sivers:1970tr}.

Over the past decade, the bootstrap approach has been revived with modern tools. Positivity bounds
on effective field theories~\cite{Adams:2006sv,Arkani-Hamed:2020blm,Caron-Huot:2020cmc} and numerical
explorations of the space of consistent
amplitudes~\cite{Paulos:2017fhb,Guerrieri:2021ivu,Huang:2020nqy,Albert:2024yap}, including
supersymmetric settings~\cite{Berman:2023jys,Berman:2024wyt}, have located string amplitudes at or
near special points of the allowed regions; see Ref.~\cite{Mizera:2023tfe} for a review. A
complementary line of work asks whether string amplitudes are singled out analytically. An argument
for the asymptotic uniqueness of the Veneziano amplitude was given in Ref.~\cite{Caron-Huot:2016icg},
while a large family of consistent deformations has since been
constructed~\cite{Coon:1969yw,Gross:1969db,Cheung:2022mkw,Geiser:2022icl,Maldacena:2022ckr,Cheung:2023uwn,Cheung:2023adk,Haring:2023zwu}.
Further structural constraints include the number and form of Regge
trajectories~\cite{Eckner:2024pqt,Eckner:2024ggx}, level truncation~\cite{Cheung:2024uhn,Cheung:2024obl},
multiparticle factorization~\cite{Arkani-Hamed:2023jry}, and hidden zeros of
amplitudes~\cite{Arkani-Hamed:2023swr}. Recently, Cheung, Remmen, Sciotti and
Tarquini~\cite{Cheung:2025tbr} showed that the residues of any tree-level, unitary amplitude with
Regge behaviour must vanish at prescribed values of the momentum transfer, the Regge zeros. Assuming
ultrasoft Regge behaviour and that these are the only zeros of the residues, they derived the linear
spectrum, the linear Regge trajectory, and ultimately the Veneziano and Virasoro--Shapiro amplitudes.

String scattering amplitudes are a rich subject in their own right. They are computed from
correlation functions on the string worldsheet~\cite{Koba:1969rw,Friedan:1985ge,Green:1987sp,Polchinski:1998rr},
and at tree level the open-superstring amplitudes of massless states are known in closed form for any
multiplicity~\cite{Mafra:2011nv,Mafra:2011nw}. Their low-energy limit reproduces supersymmetric
Yang--Mills theory~\cite{Green:1982sw}, and closed-string amplitudes are related to open-string ones by
the Kawai--Lewellen--Tye relations~\cite{Kawai:1985xq}. Field-theory structures such as on-shell
recursion~\cite{Boels:2008fc,Cheung:2010vn} and colour--kinematics duality~\cite{Bern:2008qj}, together
with the stringy versions \cite{Plahte:1970wy,Bjerrum-Bohr:2009ulz,Stieberger:2009hq, Stieberger:2023nol, Yuenyong:2024ebe, Stieberger:2016lng, Boels:2008fc, Boels:2010bv, Chang:2012qs, Cheung:2010vn, Srisangyingcharoen:2024qyx, Srisangyingcharoen:2024zko} have deepened the connection between gauge-theory and string
amplitudes.

Fermions entered dual resonance models through the work of Ramond~\cite{Ramond:1971gb} and of Neveu and
Schwarz~\cite{Neveu:1971rx,Neveu:1971iv}, and the resulting models were later understood to possess
spacetime supersymmetry~\cite{Gliozzi:1976qd}. Fermionic amplitudes bring in features that are absent
in the scalar case. Helicity amplitudes carry kinematic singularities associated with half-angle
factors~\cite{Jacob:1959at,Hara:1964zz,Cohen-Tannoudji:1968eoa}, resonances of different statistics
appear in different channels, and fermionic Regge trajectories are subject to additional analyticity
constraints such as MacDowell symmetry~\cite{MacDowell:1959zza,Collins:1977jy}. By contrast, the
analytic bootstrap programmes described above have focused almost exclusively on external scalars.
It is therefore natural to ask whether the Regge-zero argument of Ref.~\cite{Cheung:2025tbr} extends to
amplitudes with external fermions, and whether it singles out the corresponding superstring amplitude.

In this work we address this question for the colour-ordered four-point amplitude of a massless
scalar and a massless spin-$\frac12$ fermion, which contains fermionic resonances in the $s$-channel
and bosonic resonances in the $t$-channel. We extract a spinor structure that carries the
little-group transformation of the amplitude and bootstrap the remaining reduced amplitude. The Regge
zeros of the residues in each channel are controlled by the trajectory of the opposite channel, so the
two channels must be treated together. Under assumptions analogous to those of
Ref.~\cite{Cheung:2025tbr}, we find that both trajectories are linear, that the maximal spins grow as
$J_F(n)=n+\frac12$ and $J_B(m)=m+1$, and that duality between the two pole expansions fixes the reduced
amplitude to a Beta function, up to its normalisation and the ratio of Regge slopes. When the fermionic
and bosonic towers are taken to be degenerate in mass, the result coincides with the open superstring
amplitude.

The paper is organised as follows. In the rest of this section we present our conventions.
Sec.~\ref{sec:scalar} reviews the scalar bootstrap of Ref.~\cite{Cheung:2025tbr}. Sec.~\ref{sec:fermionic}
contains the fermionic analysis, and its relation to supersymmetric Yang--Mills theory and the open
superstring is discussed in Sec.~\ref{sec:discussion}. We conclude in Sec.~\ref{sec:conclusion}.
Appendix~\ref{app:counting} presents the counting argument that fixes the spins, and
Appendix~\ref{app:fermionic-partial-waves} contains the partial-wave analysis and the constraints from
unitarity.

\paragraph*{Conventions}
Throughout this work we use the mostly-plus metric signature $\eta=\text{diag}(-,+,+,+)$, so that
$p_i^2=-m_i^2$ on shell, following Ref.~\cite{Cheung:2025tbr}. We consider four-point scattering with
incoming momenta $p_1$ and $p_2$ and outgoing momenta $p_3$ and $p_4$, satisfying $p_1+p_2-p_3-p_4=0$. The Mandelstam variables are defined as
\begin{equation}
s=-(p_1+p_2)^2,\qquad t=-(p_2-p_3)^2,\qquad u=-(p_1-p_3)^2
\end{equation}
such that $s+t+u=\sum_i m_i^2$. In the $s$-channel physical region the scattering
angle satisfies $\cos\theta_s=1+2t/s$.

Consider the colour-ordered four-point amplitude $A(s,t)$ of identical scalars at tree level. Planar crossing symmetry requires $A(s,t)=A(t,s)$. The amplitude is taken to be meromorphic, with simple
poles at $s,t=\mu(n)$ for an integer level $n\ge0$. Without loss of generality, $\mu(n)$ is
monotonically increasing. The residue at level $n$,
\begin{equation}
R(n,t)=\lim_{s\to\mu(n)}\big(\mu(n)-s\big)A(s,t),
\label{eq:sv-residue}
\end{equation}
is a polynomial in $t$ whose degree $J(n)$ is the maximal spin exchanged at that level. An infinite
tower of spins on a single narrow resonance would correspond to a propagating mode of unbounded
spatial extent~\cite{Bern:2021ppb,Caron-Huot:2020cmc,Huang:2022mdb}, so $J(n)$ is required to be finite at finite $n$. Unitarity requires the partial-wave decomposition
\begin{equation}
R(n,t)=\sum_{\ell=0}^{J(n)}a_{n,\ell}\,G^{(D)}_\ell\!\left(1+\frac{2t}{\mu(n)}\right),\qquad a_{n,\ell}\ge0,
\label{eq:sv-unitarity}
\end{equation}
where $G^{(D)}_\ell$ are the $D$-dimensional Gegenbauer polynomials.

In the Regge limit, $s\to\infty$ at fixed $t$, the amplitude behaves asymptotically as
\begin{equation}
A(s,t)\sim f(t)\,(-s)^{\alpha(t)},
\label{eq:sv-regge}
\end{equation}
where $\alpha(t)$ is the leading Regge trajectory~\cite{Regge:1959mz,Collins:1977jy}. Eq.~\eqref{eq:sv-regge} is understood as an
asymptotic statement valid throughout the complex $s$-plane except on the positive real axis, where the branch cut is located. The prefactor $f(t)$ is analytic apart from simple poles
at $t=\mu(n)$ whose residues are polynomials in $s$ of degree $J(n)$, which implies the Chew--Frautschi
relation~\cite{Chew:1961ev,Chew:1962eu}
\begin{equation}
\alpha\big(\mu(n)\big)=J(n),\qquad n\ge0 .
\label{eq:sv-chew-frautschi}
\end{equation}

\section{Bootstrapping Scalar Veneziano Amplitudes}\label{sec:scalar}

This section reviews the work of Cheung \emph{et al.}~\cite{Cheung:2025tbr}, who argued that the Veneziano~\cite{Veneziano:1968yb} and Virasoro--Shapiro~\cite{Virasoro:1969me,Shapiro:1970gy} amplitudes follow uniquely from a small set of assumptions about tree-level scattering. The key observation is that the Regge behaviour of any such amplitude forces its residues to vanish at prescribed values of the momentum transfer. We concentrate on the planar four-point case, which yields the Veneziano amplitude.

In addition to the basic properties of tree-level amplitudes, namely crossing symmetry, Lorentz invariance and unitarity, the authors introduced two assumptions that single out the Veneziano amplitude:
\begin{enumerate}
\item[(i)] \emph{Ultrasoftness.} The trajectory $\alpha(t)$ is real and bijective for $t<0$, and
$\alpha(t)\to-\infty$ as $t\to-\infty$. The amplitude then decays faster than any power of $s$ at
sufficiently negative fixed $t$. This property is equivalent to superpolynomial
boundedness~\cite{Haring:2023zwu} together with bijectivity. It is exhibited by the Veneziano
amplitude, but not by its $q$-deformations~\cite{Coon:1969yw,Geiser:2022icl}, whose trajectories diverge at finite $t$, nor by
quantum field theory, whose softness is bounded~\cite{Cerulus:1964cjb,Buoninfante:2023dyd}.
\item[(ii)] \emph{Minimal zeros.} Every zero of every residue $R(n,t)$ lies exactly at a Regge zero. Additional zeros are not allowed.
\end{enumerate}

\paragraph*{Regge zeros}

The central technique is the identification of the Regge zeros of the residues $R(n,t)$. It can be shown that, at large $n$, the residues must vanish whenever $\alpha(t)$ is an integer. To see this, one integrates the Regge expression~\eqref{eq:sv-regge} around a circle of large radius $s_*$, starting
immediately above the cut on the positive real axis and ending immediately below it. Up to an
overall normalisation,
\begin{equation}
\oint_{|s|=s_*}\mathrm{d}s\,A(s,t)\simeq f(t)\,\frac{s_*^{\alpha(t)+1}\sin\pi\alpha(t)}{\alpha(t)+1}
\simeq\sum_{n}^{n_*}R(n,t),
\label{eq:sv-arc}
\end{equation}
where the second equality follows from deforming the contour onto the poles enclosed by the circle,
$\mu(n)<s_*$. Subtracting two concentric arcs that enclose a single level isolates one residue:
\begin{equation}
R(n,t)\sim f(t)\,\mu'(n)\,\mu(n)^{\alpha(t)}\sin\pi\alpha(t)
\label{eq:sv-residue-asymptotics}
\end{equation}
where $\mu'(n)\equiv\mu(n)-\mu(n-1)$, assuming $\mu'(n)\ll\mu(n)$. The expression holds up to corrections that vanish as a power of $1/n$.

The factor $\sin\pi\alpha(t)$ implies that the residues at large $n$ vanish whenever $\alpha(t)$ is
an integer. Two remarks must be taken into account. First, at the $t$-channel poles, $t=\mu(n')$, the poles of $f(t)$ cancel the zeros of $\sin\pi\alpha(t)$, so the zeros must be off resonance ($t\neq \mu(n')$).
Second, for $t>0$ the argument of the Gegenbauer polynomials in Eq.~\eqref{eq:sv-unitarity} exceeds
unity. There $G^{(D)}_\ell(x)>G^{(D)}_\ell(1)>0$ for any $\ell>0$, and the positivity of the coefficients in Eq.~\eqref{eq:sv-unitarity} gives $R(n,t)>R(n,0)>0$. The
Regge zeros are therefore the solutions of
\begin{equation}
\alpha(t)+r=0,\qquad r\in\mathbb{Z},
\label{eq:sv-regge-zeros}
\end{equation}
restricted to real roots with $t<0$ and $t\neq\mu(n')$ for all $n'\ge0$.
 
Since the Regge approximation improves with increasing level $n$, a Regge zero of $R(n,t)$ is also a Regge zero of $R(n',t)$ for all $n'>n$. Softer Regge behaviour implies more Regge zeros: an ultrasoft
trajectory crosses infinitely many negative integers, whereas a trajectory bounded from below
crosses only finitely many.

\paragraph*{Bootstrapping arguments}

By assumption (ii), the residue can be written as a product of factors $\alpha(t)+r$. Choosing the minimal value $r=1$, since any finite offset can be absorbed into a shift of $\alpha(t)$, one obtains
\begin{equation}
R(n,t)=c(n)\prod_{r=1}^{N(n)}\big(\alpha(t)+r\big)^{h}
\label{eq:sv-product}
\end{equation}
where $h$ is a constant multiplicity and $c(n)$ an arbitrary normalization. Unitarity restricts the Regge zeros to $t<0$. Assumption (i) requires $\alpha(t)$ to be bijective and monotonic, which excludes logarithmic trajectories such as that of the $q$-deformed Coon amplitude~\cite{Coon:1969yw}. Taking $\alpha(t)$ to be a polynomial of degree $d$, one finds
\begin{equation}
    J(n)=d\cdot h\cdot N(n).\label{eq:sv-degree}
\end{equation}
This follows from the fact that the degree of the residue in $t$ equals the maximal spin $J(n)$.

At large $N(n)$, Eq.~\eqref{eq:sv-product} gives
\begin{equation}
    R(n,t)\propto  N(n)^{h\alpha(t)}
\end{equation}
while Eq.~\eqref{eq:sv-residue-asymptotics} gives $R(n,t)\sim\mu(n)^{\alpha(t)}$. Comparing the two expressions gives
\[
\mu(n)\sim N(n)^h\propto J(n)^h. \]
Eq.~\eqref{eq:sv-chew-frautschi} then requires
$\alpha\big(J(n)^h\big)\sim J(n)$. Since $\alpha(t)$ is a polynomial, this forces $h=1$ and $d=1$. This gives a linear mass spectrum and a linear Regge trajectory
\begin{equation}
\mu(n)\propto J(n),\qquad \alpha(t)\propto t .
\label{eq:sv-linear}
\end{equation}

With $J(n)=N(n)$ from Eq.~\eqref{eq:sv-degree}, one can further deduce that $N(n)=n$ by establishing lower and upper bounds on $J(n)$. A detailed argument can be found in Appendix~A of Ref.~\cite{Cheung:2025tbr}. For the reader's convenience, we present a similar argument in Appendix~\ref{app:counting}.

Since the Regge trajectory is linear, we write
\begin{equation}
\alpha(t)=\alpha't+\alpha_0,
\end{equation}
where $\alpha'>0$ is required by ultrasoftness. We adopt the natural unit convention $\alpha'=1$, which absorbs the fundamental dimensionful scale into the definition of the Mandelstam invariants. The intercept $\alpha_0$ is related to the physical input $\mu(0)$. For a theory with a massless ground state, we set $\alpha_0=0$. The residue then takes the form
\begin{equation}
R(n,t)=c(n)\prod_{r=1}^{n}\big(t+r\big)=c(n)(t+1)_n
\label{eq:sv-residue-final}
\end{equation}
where $(x)_n$ denotes the rising factorial, also called the Pochhammer symbol.

\paragraph*{Level truncation}
To determine the coefficients $c(n)$, we evaluate the amplitude at $s_j=-j$ and $t_k=-k$, where $j,k\in\mathbb{Z}^+$. The amplitude can then be expressed as an $s$-channel resonance expansion:
\begin{equation}
A(s_j,t_k)=\sum_{n=0}^{k-1}\frac{c(n)(1-k)_n}{n+j}.
\label{eq:sv-truncate}
\end{equation}
Here, we have used $\mu(n)=n$. The sum terminates at $n=k-1$ because $R(n,t_k)=0$ for all $n\geq k$. This property is known as level truncation~\cite{Cheung:2024uhn,Cheung:2024obl}.

By crossing symmetry, the amplitude must also admit an equivalent $t$-channel representation:
\begin{equation}
A(s_j,t_k)=A(t_k,s_j)
=\sum_{m=0}^{j-1}\frac{c(m)(1-j)_m}{m+k}.
\label{eq:sv-truncate2}
\end{equation}

Setting $j=1$, the $t$-channel expansion reduces to a single term,
\begin{equation}
A(t_k,s_1)=\frac{c(0)}{k}.
\end{equation}
Consequently, Eq.~\eqref{eq:sv-truncate} gives
\begin{equation}
\sum_{n=0}^{k-1}\frac{c(n)(1-k)_n}{n+1}
=\frac{c(0)}{k}.
\label{eq:sv-truncate3}
\end{equation}
Using the identity
\begin{equation}
\sum_{n=0}^{k-1}\frac{(1-k)_n}{n!(n+1)}
=\frac{1}{k}, \label{eq:identity}
\end{equation}
we find that $c(n)=c(0)/n!$ satisfies Eq.~\eqref{eq:sv-truncate3} for every positive integer $k$. Moreover, since the equation for each $k$ determines the next coefficient $c(k-1)$ in terms of the preceding ones, this solution is unique.

Finally, choosing the overall normalization $c(0)=1$, we obtain the well-known Veneziano amplitude:
\begin{equation}
A(s,t)=\frac{\Gamma(-s)\Gamma(-t)}{\Gamma(-s-t)}.
\end{equation}
Strictly speaking, this is the Z-theory amplitude for coloured scalars~\cite{Carrasco:2016ldy,Cheung:2025tbr}. Different choices of the intercept $\alpha_0$ lead to a family of amplitudes. More generally, shifts in the Regge trajectory or in the mass--level relation give rise to additional solutions, known as satellite amplitudes~\cite{Gross:1969db,Cheung:2025tbr}. For the bosonic string, $\alpha_0=1$.


\section{Bootstrapping Fermionic Veneziano Amplitudes}\label{sec:fermionic}
We now extend the analysis of Sec.~\ref{sec:scalar} to an amplitude with external fermions.
Three features distinguish this case from the scalar one. First, residues on fermionic poles
carry kinematic half-angle factors~\cite{Hara:1964zz,Cohen-Tannoudji:1968eoa} and are therefore not polynomials in the momentum transfer.
We remove these factors by extracting a spinor structure that carries the little-group
transformation of the amplitude. Second, the $s$- and $t$-channels contain resonances of
different statistics. The amplitude is therefore not symmetric under $s\leftrightarrow t$, and
it is governed by two independent Regge trajectories. Third, as a consequence, the Regge zeros of the residues in each channel are controlled by the trajectory of the other channel, and the bootstrap must be carried out in both channels simultaneously.

\paragraph*{Scattering process}

We consider the scattering of a massless scalar $\phi$ and a massless spin-$\tfrac12$ fermion
$\psi$,
\begin{equation}
\phi(p_1)+\psi(p_2)\;\longrightarrow\;\psi(p_3)+\phi(p_4),
\label{eq:f-process}
\end{equation}
 We take the colour-ordered amplitude in which poles appear only in the $s$- and $t$-channels. The two
channels contain resonances of different statistics:
\begin{itemize}
\item In the $s$-channel, $\phi\psi$ forms fermionic resonances of half-integer spin. We denote
their masses by $s=\mu_F(n)$ and the maximal spin at level $n\ge0$ by $J_F(n)$.
\item In the $t$-channel, the crossed process $\psi\bar\psi\to\phi\phi$ contains bosonic resonances of integer spin. We denote their masses by $t=\mu_B(m)$ and the maximal spin at level $m\ge0$ by $J_B(m)$.
\end{itemize}
Both mass spectra are taken to be strictly increasing. In contrast with
Sec.~\ref{sec:scalar}, crossing does not relate the amplitude to itself under $s\leftrightarrow t$, since the two channels contain states of different statistics.

For massless momenta, we use the spinor-helicity formalism~\cite{Dixon:1996wi,Elvang:2015rqa} to write the momentum bispinor as $k_{a\dot a}=\lambda_a\tilde\lambda_{\dot a}$, or equivalently, $|k\rangle[k|$. The Lorentz-invariant spinor brackets $\langle ij\rangle$ and $[ij]$ satisfy
$|\langle ij\rangle|^2=|[ij]|^2=|(k_i+k_j)^2|$. 
Throughout, we adopt the all-outgoing momentum convention used in the standard literature \cite{Elvang:2015rqa}. Thus, $k_1=-p_1$, $k_2=-p_2$, $k_3=p_3$, and $k_4=p_4$. The spinors associated with $-p$ and $p$ differ only by convention-dependent phases.
We assign the (outgoing) helicity labels
\begin{equation}
h_1=h_4=0,\qquad h_2=+\tfrac12,\qquad h_3=-\tfrac12.
\label{eq:f-helicities}
\end{equation}
Under the little-group transformation
\begin{equation}
\lambda_i\to z_i\lambda_i,\qquad \tilde\lambda_i\to z_i^{-1}\tilde\lambda_i,
\label{eq:f-littlegroup}
\end{equation}
the amplitude with these helicities transforms as $A\to z_2^{-1}z_3^{+1}A$.

\paragraph*{Spinor structure and reduced amplitude}
This allows us to decompose the amplitude as
\begin{equation}
A(s,t)=K(s,t)\,M(s,t).
\label{eq:f-AKM}
\end{equation}
where $K(s,t)$ contains the spinor structure that reproduces the little-group transformation of $A(s,t)$, while $M(s,t)$ is invariant under every little-group transformation. We call $M(s,t)$ the reduced amplitude.

We require $K(s,t)$ to carry the same little-group weight as $A(s,t)$, namely $z_2^{-1}z_3^{+1}$. It must therefore contain one additional
$|3\rangle$ and one additional $|2]$ from the external fermions. The simplest Lorentz-invariant object with these properties is
\begin{equation}
K(s,t)=\langle3|k_1|2]=\langle31\rangle[12].
\label{eq:f-K}
\end{equation}
Momentum conservation, $\sum_i|i\rangle[i|=0$, implies $\langle3|k_4|2]=-K$, so this structure
is unique. Its magnitude is
\begin{equation}
|K|^2=|s|\,|u|,\qquad |K|=s\cos\tfrac{\theta}{2},
\label{eq:f-Kmag}
\end{equation}
where the second expression holds in the $s$-channel physical region. At four points, little-group-invariant Lorentz scalars are parity-even, because the parity-odd contraction $\epsilon_{\mu\nu\rho\sigma}k_1^\mu k_2^\nu k_3^\rho k_4^\sigma$ vanishes by momentum conservation. Any alternative structure with the same little-group weight is therefore $K$ times a polynomial in $s$ and $t$; for example, $\langle3|k_1k_4k_1|2]=-t\,K$. We therefore take $K$ as given in Eq.~\eqref{eq:f-K}. The reduced amplitude $M(s,t)$ is then a meromorphic function containing only the dynamical poles, whose form we now determine.

We define the residues of the reduced amplitude in the two channels by
\begin{align}
R_F(n,t)&=\lim_{s\to\mu_F(n)}\big(\mu_F(n)-s\big)M(s,t),\label{eq:f-RF}\\
R_B(m,s)&=\lim_{t\to\mu_B(m)}\big(\mu_B(m)-t\big)M(s,t).\label{eq:f-RB}
\end{align}
The partial-wave analysis of Appendix~\ref{app:fermionic-partial-waves} shows that both are polynomials, with
degrees
\begin{equation}
\begin{aligned}
\deg_tR_F(n,t)&=J_F(n)-\tfrac12,\\
\deg_sR_B(m,s)&=J_B(m)-1.
\end{aligned}
\label{eq:f-degrees}
\end{equation}
This implies that $J_B(m)$ and $J_F(n)$ are integer and half-integer, respectively.

\paragraph*{Regge zeros}
In the limit $s\to\infty$ at fixed $t$, the amplitude behaves as 
\begin{equation}
A\sim s^{\alpha_B(t)}, \label{Regge bosonic}    
\end{equation}
where $\alpha_B(t)$ is the leading bosonic trajectory. Similarly, in the limit $t\to\infty$ at fixed $s$, it behaves as
\begin{equation}
    A\sim t^{\alpha_F(s)}, \label{Regge fermionic} 
\end{equation} 
where $\alpha_F(s)$ is the leading fermionic trajectory. From
Eq.~\eqref{eq:f-Kmag}, using $u=-s-t$,
\begin{equation}
|K|\simeq
\begin{cases}
|s|, & s\to\infty,\ t\ \text{fixed},\\[2pt]
\sqrt{|s|\,|t|}, & t\to\infty,\ s\ \text{fixed}.
\end{cases}
\label{eq:f-Kasym}
\end{equation}
The reduced amplitude then obeys
\begin{align}
M(s,t)&\sim f_B(t)\,(-s)^{\hat\alpha_B(t)},\qquad &&s\to\infty,\label{eq:f-regge-s}\\
M(s,t)&\sim f_F(s)\,(-t)^{\hat\alpha_F(s)},\qquad &&t\to\infty,\label{eq:f-regge-t}
\end{align}
with the \emph{reduced trajectories}
\begin{equation}
\hat\alpha_B(t)\equiv\alpha_B(t)-1,\qquad \hat\alpha_F(s)\equiv\alpha_F(s)-\tfrac12 .
\label{eq:f-reduced-traj}
\end{equation}
The factor $\sqrt{|s|}$ in the second line of Eq.~\eqref{eq:f-Kasym} has been absorbed into $f_F(s)$.

As in Eq.~\eqref{eq:sv-chew-frautschi}, each trajectory equals the maximal spin of its own channel at the resonances:
\begin{equation}
\alpha_B(\mu_B(m))=J_B(m) \quad \text{and} \quad \alpha_F(\mu_F(n))=J_F(n).    
\end{equation} 
The reduced trajectories then become
\begin{equation}
\begin{aligned}
\hat\alpha_B\big(\mu_B(m)\big)&=J_B(m)-1=\deg_sR_B(m,s),\\
\hat\alpha_F\big(\mu_F(n)\big)&=J_F(n)-\tfrac12=\deg_tR_F(n,t).
\end{aligned}
\label{eq:f-chew-frautschi}
\end{equation}

We can determine Regge zeros of the residue $R_F(n,t)$ using the same procedure as in Sec.~\ref{sec:scalar}. Integrating Eq.~\eqref{eq:f-regge-s} along a large circle in the $s$-plane, and taking the difference of concentric arcs gives
\begin{equation}
R_F(n,t)\sim f_B(t)\,\mu_F'(n)\,\mu_F(n)^{\hat\alpha_B(t)}\sin\pi\hat\alpha_B(t),
\label{eq:f-RF-asym}
\end{equation}
with $\mu_F'(n)\equiv\mu_F(n)-\mu_F(n-1)$. The fermionic residues therefore vanish at large level on the Regge zeros:
\begin{equation}
\hat\alpha_B(t)+r=0,\qquad r\in\mathbb Z,
\label{eq:f-RZ-F}
\end{equation}
restricted to real roots with $t<0$ and $t\neq\mu_B(m)$. The restriction to $t<0$ follows from
unitarity (Appendix~\ref{app:fermionic-partial-waves}). Note that the zeros of the fermionic residues are controlled by
the bosonic trajectory, and are therefore not shifted by a half-integer.

Applying the same argument to Eq.~\eqref{eq:f-regge-t} in the $t$-plane gives
\begin{equation}
R_B(m,s)\sim f_F(s)\,\mu_B'(m)\,\mu_B(m)^{\hat\alpha_F(s)}\sin\pi\hat\alpha_F(s),
\label{eq:f-RB-asym}
\end{equation}
Accordingly, the Regge zeros are
\begin{equation}
\hat\alpha_F(s)+r=0,\qquad r\in\mathbb Z,
\label{eq:f-RZ-B}
\end{equation}
with $s\neq\mu_F(n)$. Unlike the fermionic Regge zeros, the Regge zeros~\eqref{eq:f-RZ-B} are not confined to $s<0$ by unitarity. More details can be found in Appendix~\ref{app:fermionic-partial-waves}.

\paragraph*{Bootstrapping arguments}
We adopt the assumptions of Sec.~\ref{sec:scalar} for both trajectories:
\begin{enumerate}
\item[(i)] \emph{Ultrasoftness.} The reduced trajectories $\hat\alpha_B(t)$ and $\hat\alpha_F(s)$
are real and bijective on the negative real axis, and tend to $-\infty$ as $t\to-\infty$ and $s\to-\infty$, respectively.
\item[(ii)] \emph{Minimal zeros.} Every zero of $R_F(n,t)$ and of $R_B(m,s)$ lies exactly at a
Regge zero, given by Eqs.~\eqref{eq:f-RZ-F} and~\eqref{eq:f-RZ-B}, respectively.
\end{enumerate}
As in the scalar case, assumption (ii) implies a nested structure of the residues, such that the ratios of residues at adjacent levels are polynomials. Assumption (i) excludes
logarithmic trajectories. Both reduced trajectories are therefore polynomials, of degrees
$d_B=\deg\hat\alpha_B$ and $d_F=\deg\hat\alpha_F$.

Assumption (ii) implies that each residue is a product of the factors appearing in its Regge
zeros. Choosing the minimal value $r=1$, as in Eq.~\eqref{eq:sv-product},
\begin{align}
R_F(n,t)&=c_F(n)\prod_{r=1}^{N_F(n)}\big(\hat\alpha_B(t)+r\big)^{h_F},\label{eq:f-prod-F}\\
R_B(m,s)&=c_B(m)\prod_{r=1}^{N_B(m)}\big(\hat\alpha_F(s)+r\big)^{h_B},\label{eq:f-prod-B}
\end{align}
where $h_F$ and $h_B$ are constant multiplicities and $c_F(n)$, $c_B(m)$ are normalization factors. Interestingly, the bosonic trajectory determines the fermionic residues, while the fermionic trajectory determines the bosonic residues. Matching the polynomial degrees with Eq.~\eqref{eq:f-degrees} gives
\begin{equation}
\begin{aligned}
J_F(n)-\tfrac12&=h_F\,d_B\,N_F(n),\\
J_B(m)-1&=h_B\,d_F\,N_B(m).
\end{aligned}
\label{eq:f-degree-count}
\end{equation}

Let us first consider the fermionic sector.
At large $N_F$, Eq.~\eqref{eq:f-prod-F} gives $R_F(n,t)\propto N_F(n)^{h_F\hat\alpha_B(t)}$,
while Eq.~\eqref{eq:f-RF-asym} gives $R_F(n,t)\sim\mu_F(n)^{\hat\alpha_B(t)}$. Comparing the two expressions, one obtains
\begin{equation}
\mu_F(n)\sim N_F(n)^{h_F}\propto\big(J_F(n)-\tfrac12\big)^{h_F}.
\label{eq:f-muF}
\end{equation}
Eq.~\eqref{eq:f-chew-frautschi} then requires
$\hat\alpha_F\big((J_F-\tfrac12)^{h_F}\big)\sim J_F-\tfrac12$. Since $\hat\alpha_F$ is a
polynomial of degree $d_F$, this forces $h_F\,d_F=1$, and hence $h_F=d_F=1$. Applying the same argument to the bosonic sector gives $h_B=d_B=1$ as well.
Both trajectories are linear:
\begin{equation}
\begin{aligned}
\hat\alpha_F(s)&=\alpha_F' s+\hat\alpha_F(0),\\
\hat\alpha_B(t)&=\alpha_B' t+\hat\alpha_B(0).
\end{aligned}
\label{eq:f-linear}
\end{equation}

With linear trajectories and $h_F=h_B=1$, the counting argument of Appendix~\ref{app:counting}
applies to each channel. This gives
\begin{equation}
J_F(n)=n+\tfrac12,\qquad J_B(m)=m+1,\qquad n,m\ge0.
\label{eq:f-spins}
\end{equation}

We can further fix one of the slopes, say
$\alpha_B'=1$, by a choice of units. For massless ground states,
$\mu_F(0)=\mu_B(0)=0$, the intercepts are
\begin{equation}
\hat\alpha_F(0)=\hat\alpha_B(0)=0,
\label{eq:f-intercepts}
\end{equation}
so that $\hat\alpha_F(s)=\alpha_F' s$ and $\hat\alpha_B(t)=t$. The residues then take the form
\begin{equation}
\begin{aligned}
R_F(n,t)&=c_F(n)\,(t+1)_n,\\
R_B(m,s)&=c_B(m)\,(\alpha_F' s+1)_m,
\end{aligned}
\label{eq:f-residues}
\end{equation}
with spectra $\mu_F(n)=n/\alpha_F'$ and $\mu_B(m)=m$.

\paragraph*{Level truncation and duality}
What remains is to determine the normalisation factors $c_F(n)$ and $c_B(m)$. In Sec.~\ref{sec:scalar} this
was achieved by crossing symmetry, $A(s,t)=A(t,s)$. However, the two channels contain different resonances, so no such symmetry exists. Its role is taken over by duality: the reduced amplitude must be reproduced both by its sum over fermionic poles and by its sum over bosonic
poles.

Consider the points
\begin{equation}
t_k=-k,\qquad s_j=-\frac{j}{\alpha_F'},\qquad j,k\in\mathbb Z^+,
\label{eq:f-points}
\end{equation}
at which $\hat\alpha_B(t_k)=-k$ and $\hat\alpha_F(s_j)=-j$. By Eq.~\eqref{eq:f-residues},
$R_F(n,t_k)=0$ for all $n\ge k$. One can write $M(s_j,t_k)$ in the fermionic-pole expansion as
\begin{equation}
M(s_j,t_k)=\alpha_F'\sum_{n=0}^{k-1}\frac{c_F(n)\,(1-k)_n}{n+j}.
\label{eq:f-sexp}
\end{equation}
Likewise, $R_B(m,s_j)=0$ for all $m\ge j$, and the bosonic-pole expansion at $t=t_k$ gives
\begin{equation}
M(s_j,t_k)=\sum_{m=0}^{j-1}\frac{c_B(m)\,(1-j)_m}{m+k}.
\label{eq:f-texp}
\end{equation}
Equating Eqs.~\eqref{eq:f-sexp} and~\eqref{eq:f-texp} for all $j,k\in\mathbb Z^+$ yields
\begin{equation}
\alpha_F'\sum_{n=0}^{k-1}\frac{c_F(n)(1-k)_n}{n+j}
=\sum_{m=0}^{j-1}\frac{c_B(m)(1-j)_m}{m+k}.
\label{eq:f-duality}
\end{equation}
Setting $(j,k)=(1,1)$ gives $\alpha_F' c_F(0)=c_B(0)$. For $j=1$ and arbitrary $k$, the right-hand side reduces to $c_B(0)/k$. Using the identity~\eqref{eq:identity}, one obtains
\begin{equation}
    \alpha_F' c_F(n)=\frac{c_B(0)}{n!}.
\end{equation}
Similarly, the equations with
$k=1$ determine the $c_B(m)$ in the same way, giving $c_B(m)=c_B(0)/m!$. 
Consequently, the coefficients can be written in terms of the constant $C\equiv c_B(0)$:
\begin{equation}
c_F(n)=\frac{C}{\alpha_F'\,n!},\qquad c_B(m)=\frac{C}{m!}.
\label{eq:f-coefficients}
\end{equation}

Summing the poles with the coefficients~\eqref{eq:f-coefficients} gives the reduced amplitude
\begin{equation}
M(s,t)=C\,\frac{\Gamma(-\alpha_F' s)\,\Gamma(-t)}{\Gamma(-\alpha_F' s-t)},
\label{eq:f-M}
\end{equation}
and hence the full amplitude reads
\begin{equation}
A(s,t)=CK(s,t)\,\frac{\Gamma(-\alpha_F' s)\,\Gamma(-t)}{\Gamma(-\alpha_F' s-t)},
\label{eq:f-A}
\end{equation}
where $K(s,t)=\langle 3|k_1|2]$ for the considered scattering process. One can verify directly that Eq.~\eqref{eq:f-M} has poles at $s=n/\alpha_F'$ and $t=m$. Its Regge limits are
$M\sim\Gamma(-t)(-\alpha_F' s)^{t}$ and $M\sim\Gamma(-\alpha_F' s)(-t)^{\alpha_F' s}$, in agreement
with Eqs.~\eqref{eq:f-regge-s} and~\eqref{eq:f-regge-t}. The amplitude is ultrasoft in both
channels for any $\alpha_F'>0$.

\section{Discussion}\label{sec:discussion}
We have obtained the fermionic amplitude~\eqref{eq:f-A} from the two assumptions stated above. The reduced amplitude is uniquely determined up to the overall normalisation and the ratio of slopes $\alpha_F'$. Several aspects of the derivation differ from the scalar case and deserve further study.

First, the half-integer spin of the fermionic resonances enters only through the spinor structure $K$ and the spectrum relation~\eqref{eq:f-chew-frautschi}. Factoring out $K(s,t)$ yields the reduced amplitude $M(s,t)$, whose residues are the usual polynomials of integer degree.

Second, one of the slopes can be fixed by a choice of units for the Mandelstam variables; our choice is $\alpha'_B=1$. This leaves $\alpha'_F$ undetermined. Equal slopes,
$\alpha_F'=\alpha_B'$, require an additional assumption. A natural choice is that the
fermionic and bosonic towers are degenerate in mass, $\mu_F(n)=\mu_B(n)$ for all $n\ge0$. Such a degeneracy between bosonic and fermionic states is the hallmark of unbroken spacetime supersymmetry. With this further assumption, the bootstrapped amplitude reproduces the open superstring amplitude~\cite{Green:1982sw,Green:1987sp,Polchinski:1998rr,Mafra:2011nv}
\begin{equation}
A_{\text{superstring}}=A_{\text{SYM}}\frac{\Gamma(1-s)\Gamma(1-t)}{\Gamma(1-s-t)} \label{superstring}
\end{equation}
where $A_{\text{SYM}}\sim K(s,t)\frac{s+t}{st}=K(s,t)\big(\frac{1}{s}+\frac{1}{t} \big)$.

We now show that the supersymmetric Yang--Mills amplitude~\cite{Brink:1976bc} $A_{\text{SYM}}$ has the desired form. At tree level, the $s$-channel contribution involves
\begin{equation}
\langle3|(k\!\!\!/_1+k\!\!\!/_2)|2]=\langle3|k_1|2]=K(s,t).
\label{eq:f-numerators}
\end{equation}
In the $t$-channel, the fermion current $\langle3|\gamma^\mu|2]$ is contracted with the scalar current
$(k_1-k_4)_\mu$, giving $2K(s,t)$ by momentum conservation. Accordingly,
\begin{equation}
A_{\text{SYM}}= K(s,t)\big(\frac{g_Y^2}{s}+\frac{2g^2}{t}\big),    
\end{equation}
where $g_Y$ and $g$ denote the Yukawa and gauge
couplings for canonically normalised fields. Supersymmetry fixes the Yukawa coupling in terms of the gauge coupling as $g_Y=\sqrt{2}g$~\cite{Martin:1997ns}, so that $A_{\text{SYM}}=2g^2K(s,t)\big(\frac1s+\frac1t\big)$. Substituting this into Eq.~\eqref{superstring} and using $\Gamma(1-x)=-x\,\Gamma(-x)$ reproduces Eq.~\eqref{eq:f-A} with $\alpha_F'=1$ and $C=-2g^2$.

\section{Conclusion}\label{sec:conclusion}

We have extended the Regge-zero bootstrap of Ref.~\cite{Cheung:2025tbr} from identical scalars to the
scattering of a massless scalar and a massless spin-$\frac12$ fermion. Factoring the colour-ordered
amplitude as $A=KM$, with $K=\langle3|k_1|2]$ carrying the little-group weight and the half-angle
factors, reduces the problem to a little-group-invariant amplitude $M(s,t)$ whose residues are
polynomials of integer degree, $J_F(n)-\frac12$ and $J_B(m)-1$, in the two channels. In contrast with
the scalar case, the $s$- and $t$-channels contain resonances of different statistics and are governed
by two trajectories. The residues on the fermionic poles vanish at Regge zeros controlled by the bosonic
trajectory, and vice versa.

Assuming ultrasoft Regge behaviour and the minimal zeros, we found that
both reduced trajectories are linear and that the Regge zeros are simple roots. The counting
argument of Appendix~\ref{app:counting} then fixes the maximal spins to $J_F(n)=n+\frac12$ and
$J_B(m)=m+1$, so that the lowest state in each channel has the smallest spin allowed by the helicity
structure. The role of crossing symmetry in the scalar case is taken over by duality between the
fermionic- and bosonic-pole expansions of $M$. On the kinematic points where both expansions
truncate, duality fixes all normalisations up to an overall constant, and the reduced amplitude is the
Beta function~\eqref{eq:f-M}. The ratio of Regge slopes is not fixed by this bootstrapping criteria. Assuming
degenerate fermionic and bosonic spectra equates the slopes, and the bootstrapped amplitude then
coincides with the open superstring amplitude, whose field-theory limit is the supersymmetric
Yang--Mills amplitude with $g_Y=\sqrt2\,g$.

Unitarity plays a different role in the two channels. The $s$-channel is elastic, and the positivity
of the partial-wave coefficients confines the fermionic Regge zeros to $t<0$, similar to the scalar case.
The $t$-channel connects a fermion pair to a scalar pair, and unitarity does not fix the sign of the
corresponding coefficients. The absence of bosonic Regge zeros at $s>0$ is therefore cannot be concluded. 

Several extensions suggest themselves. First, it would be interesting to identify the additional
consistency conditions that fix the ratio of slopes, for instance from other amplitudes that share the
same towers of states, or from higher-point amplitudes. Second, the analysis could be extended to the
non-planar amplitude, where crossing under $s\leftrightarrow u$ enters, in analogy with the
Virasoro--Shapiro part of Ref.~\cite{Cheung:2025tbr}, and to other helicity configurations. Third,
fermionic partial waves are naturally analytic in $\sqrt{s}$, and Regge trajectories of opposite parity
are related by MacDowell symmetry~\cite{MacDowell:1959zza,Collins:1977jy}. Understanding how this
structure interacts with the ultrasoftness and minimal-zero assumptions, in particular for massive
external fermions, may clarify which features of the superstring spectrum are required by consistency.
Finally, the five-point analysis of Ref.~\cite{Cheung:2025tbr} and the known higher-point string
amplitudes~\cite{Mafra:2011nv,Arkani-Hamed:2024nzc} suggest that the present approach could be applied
to fermionic amplitudes with more external legs.

\begin{acknowledgments}
We are grateful to the National Science, Research and Innovation Fund (NSRF) via the Program Management Unit for Human Resources \& Institutional Development, Research and Innovation for support under grant number B39G680009.
\end{acknowledgments}

\appendix

\section{Constraint counting}\label{app:counting}
In this appendix we show that the maximal spin at each level grows by exactly one unit per level. We treat the scalar amplitude of Sec.~\ref{sec:scalar} and the two channels of the fermionic amplitude of Sec.~\ref{sec:fermionic} simultaneously. The argument establishes lower and upper bounds on the degree of the residues.

Let $M(x,y)$ denote the amplitude in the scalar case, or the reduced amplitude in the fermionic
case, where $x$ denotes the variable of the channel under consideration and $y$ the crossed variable.
The amplitude has simple poles at $x=\mu(n)$, $n\ge0$, with $\mu(n)$ strictly increasing and
residues
\begin{equation}
R(n,y)=\lim_{x\to\mu(n)}\big(\mu(n)-x\big)M(x,y).
\label{eq:A-residue}
\end{equation}
In the Regge limit of large $y$ and fixed $x$, the trajectory $\hat\alpha$ controls the Regge behaviour,
\begin{equation}
M(x,y)\sim f(x)\,(-y)^{\hat\alpha(x)},\qquad y\to\infty .
\label{eq:A-regge}
\end{equation}
The trajectory $\hat\alpha(x)$ evaluated at a pole $x=\mu(n)$ equals the degree of the residue.
Conversely, the behaviour at large $x$ and fixed $y$, $M(x,y)\sim g(y)\,(-x)^{\beta(y)}$, is controlled by
the trajectory $\beta$ of the crossed channel, which determines the Regge zeros of $R(n,y)$. In the scalar
case $\beta=\alpha$, while in the fermionic amplitude $\beta$ is the reduced trajectory of the other
channel.

The three cases are distinguished by a constant offset $a_i$ in the master relation
\begin{equation}
\hat\alpha\big(\mu(n)\big)=J_i(n)+a_i\equiv D_i(n),\qquad i\in\{S,F,B\},
\label{eq:A-master}
\end{equation}
where $J_i(n)$ is the maximal spin at level $n$ and $D_i(n)$ is the degree of $R(n,y)$ in $y$. The
offset originates in the kinematic half-angle factors removed by the spinor structure $K$. The
assignments are summarised in Table~\ref{tab:A-cases}.
 
\begin{table}[h]
\centering
\renewcommand{\arraystretch}{1.3}
\begin{tabular}{@{}lcccccc@{}}
\hline\hline
Case & $x$ & $y$ & $\hat\alpha$ & $\beta$ & $a_i$ & $D_i$\\
\hline
$S$ & $s$ & $t$ & $\alpha$ & $\alpha$ & $0$ & $J$\\
$F$ & $s$ & $t$ & $\alpha_F-\tfrac12$ & $\alpha_B-1$ & $-\tfrac12$ & $J_F-\tfrac12$\\
$B$ & $t$ & $s$ & $\alpha_B-1$ & $\alpha_F-\tfrac12$ & $-1$ & $J_B-1$\\
\hline\hline
\end{tabular}
\caption{The three cases of the counting argument.}
\label{tab:A-cases}
\end{table}

\paragraph*{Lower bound.}
Since $\mu(n)$ is strictly increasing and $\hat\alpha$ is linear with $\alpha'>0$,
Eq.~\eqref{eq:A-master} implies that $D_i(n)$ is strictly increasing. Being the degree of a
polynomial, $D_i(n)$ is also a non-negative integer. Hence $D_i(n+1)\ge D_i(n)+1$, and
\begin{equation}
D_i(n)\ge D_i(0)+n\ge n .
\label{eq:A-lower}
\end{equation}

\paragraph*{Upper bound.}
Let us choose a level $n_*\geq1$ and a Regge zero $y_*$ of the residues such that
\begin{equation}
    \beta(y_*)+D_i(n_*)=0. \label{eq:upper1}
\end{equation}
Since the residues are products of the factors $\beta(y)+r$ with $r=1,\dots,D_i(n)$, all residues
$R(n,y_*)$ with $n\geq n_*$ vanish. We then compute a large arc integral of radius $x_*$ in the
$x$-plane,
\begin{equation}
    \mathcal{I}_r=\oint_{x_*} dx \ x^{r-1} M(x,y_*) \sim (x_*)^{r-D_i(n_*)},\label{eq:upper2}
\end{equation}
which vanishes for any $1 \leq r < D_i(n_*)$.
Equivalently, the integral~\eqref{eq:upper2} can be expressed as a sum of residues,
\begin{equation}
    \mathcal{I}_r \sim \sum_{n=0}^{n_*-1}\mu(n)^{r-1}R(n,y_*)=0, \label{eq:A-sumrules}
\end{equation}
for $r$ in the above range. The summation stops at $n_*-1$, since $R(n,y_*)=0$ for $n\geq n_*$.

Eqs.~\eqref{eq:A-sumrules} are $D_i(n_*)-1$ linear equations for the $n_*$ residues $R(n,y_*)$. In the marginal case $r=D_i(n_*)$, the integral~\eqref{eq:upper2} is a non-vanishing constant, which provides one further, inhomogeneous equation. Altogether, these form a set of $D_i(n_*)$ linear equations for the variables $R(n,y_*)$.

Since the coefficient matrix $\mu(n)^{r-1}$ is of Vandermonde type with distinct nodes, these equations admit a solution only if the number of constraints does not exceed the number of variables. Therefore,
\begin{equation}
    D_i(n_*) \leq n_* .\label{eq:A-upper}
\end{equation}
Combining Eq.~\eqref{eq:A-lower} at $n=1$ with Eq.~\eqref{eq:A-upper} at $n_*=1$ gives
$D_i(0)+1\le D_i(1)\le1$, hence $D_i(0)=0$. Eqs.~\eqref{eq:A-lower} and~\eqref{eq:A-upper} then
give $D_i(n)=n$ for all $n\ge0$. By the master relation~\eqref{eq:A-master},
\begin{equation}
J_i(n)=n-a_i ,
\label{eq:A-result}
\end{equation}
that is,
\begin{equation}
J(n)=n,\qquad J_F(n)=n+\tfrac12,\qquad J_B(m)=m+1,
\label{eq:A-spins}
\end{equation}
for $n,m\ge0$. In particular, the lowest state in each channel has the smallest spin compatible
with the helicity structure: spin $0$ for the scalar amplitude, spin $\tfrac12$ in the fermionic
channel, and spin $1$ in the bosonic channel.

\section{Fermionic partial-wave expansion and unitarity}
\label{app:fermionic-partial-waves}

In this appendix, we show how the partial-wave expansion of helicity
amplitudes determines the polynomial degrees of the residues of the
reduced amplitude. We then use partial-wave unitarity to constrain
the zeros of these residues. The analysis follows the standard helicity
partial-wave formalism~\cite{Jacob:1959at,Bellazzini:2022wzv}.

For a $2\to 2$ scattering process, let
\begin{equation}
    \lambda = \lambda_1-\lambda_2,
    \qquad
    \lambda' = \lambda_3-\lambda_4,
\end{equation}
denote the helicity differences of the incoming and outgoing pairs, respectively. The helicity amplitude admits the partial-wave
expansion
\begin{equation}
    \mathcal{A}_{\lambda\lambda'}(s,\theta)
    =
    \sum_J (2J+1)\,
    a^J_{\lambda\lambda'}(s)\,
    d^J_{\lambda'\lambda}(\theta),
    \label{eq:app-helicity-pw}
\end{equation}
where $a^J_{\lambda\lambda'}(s)$ are the partial-wave amplitudes
and $d^J_{\lambda'\lambda}(\theta)$ are Wigner $d$-functions.
Overall normalization factors depend on convention and will not affect
the polynomial-degree arguments below.

The Wigner functions can be written in terms of Jacobi polynomials as
\cite{Bellazzini:2022wzv}
\begin{align}
d^J_{\lambda'\lambda}(\theta)
={}&
(-1)^{\lambda'-\lambda}
\sqrt{
\frac{(J+\lambda')!(J-\lambda')!}
     {(J+\lambda)!(J-\lambda)!}
}
\left(\sin\frac{\theta}{2}\right)^{\lambda'-\lambda}
\nonumber\\
&\times
\left(\cos\frac{\theta}{2}\right)^{\lambda+\lambda'}
P_{J-\lambda'}^{(\lambda'-\lambda,\lambda+\lambda')}
(\cos\theta),
\label{eq:app-wigner-jacobi}
\end{align}
for $\lambda'\geq\lambda$. Here,
$P_n^{(a,b)}(x)$ denotes a Jacobi polynomial of degree $n$.

The two cases needed below are
\begin{equation}
    d^J_{\frac12,\frac12}(\theta)
    =
    \cos\frac{\theta}{2}\,
    P_{J-\frac12}^{(0,1)}(\cos\theta),
    \label{eq:app-d-fermion}
\end{equation}
for fermionic exchange in the $s$ channel, and
\begin{equation}
    d^J_{1,0}(\theta)
    =
    -\sqrt{\frac{J+1}{J}}\,
    \sin\frac{\theta}{2}\cos\frac{\theta}{2}\,
    P_{J-1}^{(1,1)}(\cos\theta),
    \label{eq:app-d-boson}
\end{equation}
for bosonic exchange in the crossed $t$ channel.

\paragraph*{Fermionic resonances in the $s$ channel}

On the pole $s=\mu_F(n)$, the helicity configuration
(\ref{eq:f-helicities}) has $\lambda=\lambda'=-1/2$.
Using
$d^J_{-\frac12,-\frac12}(\theta)
=d^J_{\frac12,\frac12}(\theta)$, the residue of the full amplitude is
\begin{equation}
    \lim_{s\to\mu_F(n)}
    \bigl(\mu_F(n)-s\bigr)\mathcal{A}(s,t)
    =
    \sum_J
    \gamma^F_{n,J}\,
    d^J_{\frac12,\frac12}(\theta_s),
    \label{eq:app-factorization-F}
\end{equation}
where the normalization factors, including $2J+1$, have been absorbed
into $\gamma^F_{n,J}$.

For an elastic channel, factorization and unitarity imply that the
partial-wave residue coefficients are proportional to squared
three-point couplings. We choose the convention such
that
\begin{equation}
    \gamma^F_{n,J}\geq 0.
    \label{eq:app-unitarity-F}
\end{equation}
This positivity is the relevant consequence of unitarity for the
argument below.

Using Eq.~\eqref{eq:app-d-fermion}, each spin-$J$ contribution contains
a common factor $\cos(\theta_s/2)$ multiplying a Jacobi polynomial.
The kinematic factor $K(s,t)$ 
contains the same half-angle factor at the $s$-channel pole.
Dividing by $K$ therefore removes this factor.
The reduced residue can be written as
\begin{equation}
    R_F(n,t)
    =
    \sum_{J=\frac12,\frac32,\ldots}^{J_F(n)}
    b^F_{n,J}\,
    P_{J-\frac12}^{(0,1)}
    \left(1+\frac{2t}{\mu_F(n)}\right),
    \label{eq:app-RF-expansion}
\end{equation}
where $J_F(n)$ is the maximal spin at level $n$. The coefficients
$b^F_{n,J}$ differ from $\gamma^F_{n,J}$ only by common positive
normalization factors and hence satisfy
\begin{equation}
    b^F_{n,J}\geq 0.
    \label{eq:app-bF-positivity}
\end{equation}
Consequently, the degree of $R_F(n,t)$ is 
$J_F(n)-\tfrac12$.

\paragraph*{Unitarity and zeros of the fermionic residue}

The positivity condition
\eqref{eq:app-bF-positivity} allows us to constrain the location of the zeros of $R_F(n,t)$. The Jacobi polynomials $P_\ell^{(0,1)}$ have all their zeros in the interval $(-1,1)$. Moreover, each of them is strictly increasing for $x>1$ when $\ell\geq 1$, while $P_0^{(0,1)}(x)=1$. Therefore,
\begin{equation}
    P_\ell^{(0,1)}(x)\geq P_\ell^{(0,1)}(1)=1,
\qquad x>1,\quad \ell=0,1,2,\ldots.
\end{equation}
For $t>0$, the argument
of Eq.~\eqref{eq:app-RF-expansion} is greater than one. Thus,
\begin{equation}
    R_F(n,t)\geq R_F(n,0)>0\qquad \text{for }t>0.
    \label{eq:app-no-positive-t-zeros-F}
\end{equation}
This means that the fermionic residue has
no zeros at positive $t$. The Regge zeros (\ref{eq:f-RZ-F}) are confined to $t<0$.

\paragraph*{Bosonic resonances in the crossed $t$ channel}

In the crossed channel, the scalar pair has helicity difference zero,
whereas the fermion pair has helicity difference $\pm1$. For a fixed
helicity component, the partial waves therefore involve
$d^J_{1,0}(\theta_t)$ or its helicity-conjugate counterpart.

At a bosonic pole $t=\mu_B(m)$, the residue of the full amplitude can
be expressed as
\begin{equation}
    \lim_{t\to\mu_B(m)}
    \bigl(\mu_B(m)-t\bigr)\mathcal{A}(s,t)
    =
    \sum_J
    \gamma^B_{m,J}\,
    d^J_{1,0}(\theta_t),
    \label{eq:app-factorization-B}
\end{equation}
where the normalization factors have been absorbed into
$\gamma^B_{m,J}$. These coefficients contain products of couplings
between the exchanged bosonic state and the scalar-pair and
fermion-pair channels.

Using Eq.~\eqref{eq:app-d-boson}, every spin-$J$ contribution contains
the common factor
$\sin(\theta_t/2)\cos(\theta_t/2)$, multiplying a Jacobi polynomial.
The crossed-channel expression for $K(s,t)$ contains the same
half-angle dependence. Dividing by $K$ removes this factor, leaving
a polynomial in $\cos\theta_t$. 
The reduced residue takes the form
\begin{equation}
    R_B(m,s)
    =
    \sum_{J=1}^{J_B(m)}
    b^B_{m,J}\,
    P_{J-1}^{(1,1)}
    \left(1+\frac{2s}{\mu_B(m)}\right),
    \label{eq:app-RB-expansion}
\end{equation}
where $J_B(m)$ is the maximal integer spin at level $m$. Its polynomial degree
is $J_B(m)-1$.

\paragraph*{Unitarity and zeros of the bosonic residue}

In contrast to the $s$ channel, the crossed channel is not elastic. The residue
\eqref{eq:app-factorization-B} connects the fermion pair $\psi\bar\psi$ to the scalar pair
$\phi\phi$, so that factorization gives
\begin{equation}
    \gamma^B_{m,J}=\sum_X g_{\psi\bar\psi X}\,g_{X\phi\phi},
    \label{eq:app-gammaB}
\end{equation}
where the sum runs over the spin-$J$ states $X$ at level $m$. Each coefficient is a product of
two different couplings rather than a squared coupling. Therefore, unitarity does not fix the
sign of $\gamma^B_{m,J}$, nor that of the coefficients $b^B_{m,J}$ in
Eq.~\eqref{eq:app-RB-expansion}.

For $s>0$, the argument of Eq.~\eqref{eq:app-RB-expansion} is greater than one, where each
Jacobi polynomial $P^{(1,1)}_{J-1}$ is positive. Without a definite sign for the coefficients,
however, their sum may still vanish. This means that unitarity alone does not exclude zeros of $R_B(m,s)$ at positive
$s$, and the Regge zeros \eqref{eq:f-RZ-B} are not confined to $s<0$ by this argument.

\bibliography{apssamp}

\end{document}